\documentstyle[prl,aps,multicol,epsfig,rotating]{revtex}

\renewcommand{\narrowtext}{\begin{multicols}{2}}
\renewcommand{\widetext}{\end{multicols}}

\begin{document}
\draft

\newcommand{\lsim}   {\mathrel{\mathop{\kern 0pt \rlap
  {\raise.2ex\hbox{$<$}}}
  \lower.9ex\hbox{\kern-.190em $\sim$}}}
\newcommand{\gsim}   {\mathrel{\mathop{\kern 0pt \rlap
  {\raise.2ex\hbox{$>$}}}
  \lower.9ex\hbox{\kern-.190em $\sim$}}}
\def\beq{\begin{equation}}
\def\eeq{\end{equation}}
\def\ba{\begin{eqnarray}}
\def\ea{\end{eqnarray}}
\def\d{{\rm d}}

\newcommand{\promille}{%
    \relax\ifmmode\promillezeichen
          \else\leavevmode\(\mathsurround=0pt\promillezeichen\)\fi}
  \newcommand{\promillezeichen}{%
    \kern-.05em%
    \raise.5ex\hbox{\the\scriptfont0 0}%
    \kern-.15em/\kern-.15em%
    \lower.25ex\hbox{\the\scriptfont0 00}}

\def\ap{\approx}
\def\eff{{\rm eff}}
\def\P{{\mathcal P}}
\def\theta{\vartheta}
\def\epsilon{\varepsilon}

\def\tmin{t_{\rm min}}
\def\zmin{z_{\rm min}}
\def\zmax{z_{\rm max}}


\title{Signatures of AGN model for UHECR }

\author{V. Berezinsky$^{1)}$, A. Gazizov$^{2)}$ and S. Grigorieva$^{3)}$}

\address{$^{1)}$INFN, Laboratori Nazionali del Gran Sasso, I-67010 Assergi
(AQ), Italy \\
$^{2)}$ B.I.Stepanov Institute of Physics of the National Academy of
Sciences of Belarus,
F.Skariny Ave. 68, 220062 Minsk, Belarus\\
$^{3)}$ Institute for Nuclear Research of Russian Academy of Sciences,
60th Anniversary of October Revolution Prospect 7A, 117312 Moscow, Russia}
\date{\today}

\maketitle
\begin{abstract}
We demonstrate that the energy spectra of Ultra High Energy Cosmic
rays (UHECR) as observed by AGASA, Fly's Eye, HiRes and Yakutsk 
detectors, have the imprints of UHE proton interaction with the CMB 
radiation in the form of the dip at $E\sim 1\times 10^{19}$~ eV,
of the beginning of the GZK cutoff, and of very good agreement with
calculated spectrum shape. We argue that these data, combined
with small-angle clustering and correlation with AGN (BL Lacs),    
point to the AGN model of UHECR origin at energies 
$E \lsim 1\times 10^{20}$~eV. Our consideration includes also the case when  
correlation with BL Lacs is excluded from the analysis. The excess 
of the events at $E \gsim 1\times 10^{20}$~eV , which is observed by 
AGASA (but absent in HiRes data) can be explained by another component
of UHECR, e.g. by UHECR from superheavy dark matter.
\end{abstract}
\pacs{PACS numbers: 98.70.Sa, 98.54.Cm, 95.85.S}

\narrowtext

The nature of signal carriers and of the sources of UHECR are not yet 
established (for recent reviews see \cite{NaWa,BhSi,Olinto}). The most 
natural primary particles are extragalactic protons.
Due to interaction with the CMB radiation the  Ultra High Energy (UHE) protons
from extragalactic sources are predicted to have a sharp
steepening of energy spectrum, so called GZK cutoff \cite{GZK}.
For uniformly distributed sources, the GZK
cutoff is characterized by energy $E_{1/2}$ where the integral
spectrum calculated with energy losses taken into account becomes
twice lower than the power-law extrapolation from low energies, 
$E_{1/2} = 5.3\times 10^{19}$~eV \cite{BG}.

The particles with energies higher than $1\times 10^{20}$~eV are
undoubtly observed. There are at least two ``golden'' events at
energies $2 - 3 \times 10^{20}$~eV \cite{AGASAHE,FEHE} with very
reliable energy determination (see also discussion in Ref.\cite{AGASA}). 
However, the real contradiction with the 
existence of the GZK cutoff is observed only by AGASA (see Fig.1). 
Data of HiRes \cite{HiRes} are in good agreement with presence of the GZK
cutoff (Fig. 1). The data of other detectors are not as conclusive, though a
few events with energy higher than $1\times 10^{20}$~eV are observed
there (see \cite{NaWa}) for a review and \cite{BaWa} for recent discussion). 
  
In this Letter we shall demonstrate that the observed spectra have the 
imprints of UHE proton interaction with the CMB radiation in the form of
the dip at the energy $E \sim 1\times 10^{19}$~eV, produced by 
$p+\gamma_{\rm CMB} \to p+e^++e^-$ interaction, in the form of
the beginning of the GZK cutoff,  and in the form of good 
agreement between predicted and observed spectra. We argue that at
least at energies up to $1\times 10^{20}$~eV the data (spectrum, 
small-scale clustering \cite{doubl}, and probably the 
correlation with BL Lac sources \cite{BLLac} can be explained in the
model with AGN as the sources of UHE protons.

Calculating the spectra, we shall use the cosmological parameters as 
follows from recent observations \cite{cosm}: flat universe with 
$\Omega_{\rm tot}=1$ and $\Omega_{\Lambda}=0.7$. At small redshifts 
$z$,  neutrinos, baryons and CDM behave as 
non-relativistic matter with $\Omega_m=0.3$. We shall use the Hubble 
constant $H_0=70$~km/s Mpc. The relation between time and redshift is
given by 
\beq
dt=\frac{dz}{H_0(1+z)\sqrt{\Omega_m(1+z)^3+\Omega_{\Lambda}}},
\label{dt/dz}
\eeq
The spectrum of UHE protons in the model with 
uniform distribution of the sources and with the power-law generation 
spectrum can be calculated using the formalism of Ref.\cite{BG}, 
with the continuous energy losses from Refs.\cite{BGG,StPr} (note
the difference in formulae due to cosmology with $\Lambda$ term):
\ba
J_p(E)& = &(\gamma_g-2)\frac{c}{4\pi}\frac{{\cal L}_0}{H_0}
\int_0^{z_m}\frac{dz_g(1+z_g)^{m-1}}{\sqrt{\Omega_m(1+z_g)^3+
\Omega_{\Lambda}}} \nonumber
\\ &\times &
\left [E_g(E,z_g)\right ]^{-\gamma_g} \frac{dE_g(z_g)}{dE},  
\label{flux}
\ea
where $z_m$ is a maximum redshift in the evolution of the sources, 
$z_g$ is a redshift at generation and $E_g(z_g)$ is energy of a proton
at generation, for present ($z=0$) energy $E$; 
${\cal L}_0=n_0 L_p$ is cosmic ray (CR)
emissivity at $z=0$ ($n_0$ and $L_p$ are space density of the sources and
their CR luminosity, respectively). As the general case we assume
cosmological evolution of the sources given by 
${\cal L}(z)={\cal L}_0(1+z)^m$, where the absence of evolution
corresponds to $m=0$. All energies in Eq.(\ref{flux}) are given in GeV
and luminosities in GeV/s. Dilation of energy
interval is given by \cite{BG}, modified by Eq.(\ref{dt/dz}) as
\ba
& \frac{dE_g(z_g)}{dE} = (1+z_g) \nonumber 
\\ &\times  \exp\left[\frac{1}{H_0}\int_0^{z_g}
\frac{dz(1+z)^2}{\sqrt{\Omega_m(1+z)^3+\Omega_{\Lambda}}} 
\left( \frac{db_0(E^{\prime})}{dE^{\prime}}\right) \right] 
\label{dilation}  
\ea
where $b_0(E)=dE/dt$ is the energy loss due to interaction with CMB photons at 
$z=0$. 
Derivative $db_0(E')/dE'$ at $z=0$ ( given in Ref.\cite{BGG}) is taken 
at energy $E'= (1+z)E_g(E,z)$.

For particles with energies $E\gtrsim 1\times 10^{17}$~eV, the maximum
redshift  for evolution of CR sources $z_m$ is not important
if it is larger than 4. 
In Fig.1 the calculated spectra are compared with the data of AGASA, 
HiRes, Fly's Eye and Yakutsk detectors. Note that the theoretical
spectra are the same in all four panels. We
assume the generation spectrum $\propto E^{-2}$ at $E \leq E_c$ and 
$\propto E^{-\gamma_g}$ at $E\geq E_c$ with a spectrum cutoff at 
$E_{\rm max}$. The calculations for this case are easily generalized 
from Eq.(\ref{flux}), as it is done by Eq.(10) from Ref.\cite{BGG}.
As parameters we have chosen 
$E_{\rm max}= 1\times 10^{21}$~eV, $E_c= 1\times 10^{18}$~eV, 
$\gamma_g=2.7$ for non-evolutionary ($m=0$) model and 
$\gamma_g=2.5$ for evolutionary ($m=3$) model. The required CR
emissivity for both models is 
${\cal L}_0 \approx (2.5 - 3.5)\times 10^{46}$~ergs/Mpc$^3$yr. 
The choice of $E_c$ is
motivated by the observed chemical composition. We assume that transition 
to extragalactic component occurs at the observed second knee 
\cite{FEp}-\cite{Yap}, at $E_2 \approx 4\times 10^{17}$~eV.
The observed rigidity cutoff for protons, $E/Z=(4-5) \times 10^{15}$~eV
according to KASCADE data \cite{KASCADE,Hor}, corresponds to the
cutoff energy for 
iron nuclei $ E \approx 1\times 10^{17}$~eV. Remaining gap 
$(1 - 4)\times 10^{17}$~eV can be filled 
by ultraheavy nuclei with charge up to $Z=92$ \cite{Hor}. 
The transition to the lighter chemical composition with a large fraction
of protons has been observed at $E> 3\times 10^{17}$~eV by 
AGASA \cite{AGASAp}, FE \cite{FEp}, Yakutsk \cite{Yap}   
and HiRes \cite{HiResp} detectors. 
The data of HiRes \cite{Sokol} show that at $E \gsim 6\times 10^{17}$~eV 
the protons can be the dominant component.
\vspace{5mm}  
\begin{figure}[htb]
\epsfxsize=7truecm
\centerline{\epsffile{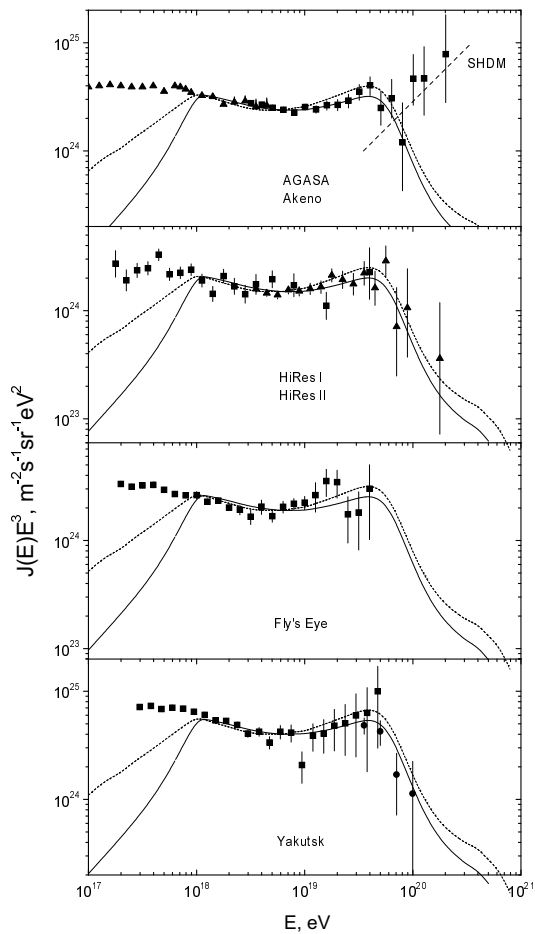}}
\vspace{1mm}
\caption{{\em The calculated spectra for non-evolutionary model 
(full lines), and evolutionary model  (dotted lines) with parameters 
indicated in the text. Both curves were first normalized to 
the AGASA data at $E=1\times 10^{18}$~eV adjusting the emissivity 
${\cal L}_0$.   To fit the data of HiRes, Fly's Eye and Yakutsk
the emissivity ${\cal L}_0$  has been scaled by factors 0.63,~0.80,~ and 1.7,
respectively. 
}}
\label{spectra}
\end{figure}
Fig.1 shows that that the signatures of interaction of UHE protons
with the CMB radiation, the  dip and the beginning of the GZK cutoff, 
are seen in the data. 

The most natural sources of the observed UHE protons are AGN. 
The required CR emissivity meet well the
local emissivity of AGN, e.g. that of Seyfert galaxies is of order 
${\cal L}_{\rm Sy} \sim n_{\rm Sy}L_{\rm Sy} \sim 1\times
10^{48}$~ergs Mpc$^3$yr. The  protons can be accelerated in AGN
up to energies of order $\sim 10^{21}$~eV \cite{acc}. An interesting 
possibility is acceleration in the jets by unipolar induction 
\cite{unip}. The correlation of UHECR with BL Lacs \cite{BLLac}, 
i.e. with AGN whose jets are directed towards us, strongly supports
this mechanism. 

Two sets of observational data favor rectilinear propagation of UHE
signal carriers in the universe. 

The first set is the above-mentioned correlation \cite{BLLac} with BL Lacs.
The second one is the small-angle clustering \cite{doubl}. Its most
natural interpretation is given \cite{stat} in terms of statistically 
occasional  arrival of two or three particles from a compact source.   
Such an interpretation needs rectilinear propagation of the primaries
and large number of sources \cite{stat}.  

Since the proton origin of UHECR is almost proved, the correlation 
with BL Lacs directly implies the rectilinear propagation of UHE protons. 
Thus, {\em these correlations  become supersensitive tools to 
measure extragalactic magnetic fields}. Below we shall discuss what 
is the scale of this field and whether this field can be already excluded.

Magnetic field must not produce the angular deflection larger than 
angular resolution of sources in the detectors, which is typically 
$\theta_{\rm res}\approx 2.5^{\circ}$. The correlation is found in
the energy range $(4 - 8)\times 10^{19}$~eV, for which the largest
attenuation length is $l_{\rm att} \sim 1000$~Mpc. The required 
upper limit for the     
magnetic field, which is homogeneous on this scale, is 
$B_l \leq 2\times 10^{-12}l_{1000}^{-1}$~G, where $l_{1000}$ is
attenuation length for $4\times 10^{19}$~eV protons in units of 1000 Mpc.
For a magnetic field with small homogeneity length  $l_{\rm hom}$ the 
required upper limit is 
\beq 
B \leq \frac{E\theta_{\rm res}}{e\sqrt{l_{\rm att}l_{\rm hom}}} \sim
6\times 10^{-10}~\mbox{G},
\label{B}
\eeq
where the numerical value is given for $l_{\rm att} \sim 1000$~Mpc 
and $l_{\hom} \sim 10$~kpc.

We argue that these fields are not excluded. 

The observed Faraday rotations give only the upper limits on large
scale extragalactic magnetic field \cite{Kro}.
All known mechanisms of generation of the large scale cosmological 
magnetic field results in extremely weak magnetic field 
$\sim 10^{-17}$~G or less (for a review see \cite{GrRu}). 
The strong magnetic field can be generated in compact 
sources, 
presumably by dynamo mechanism, and spread away by the flow of the
gas. These objects thus are surrounded by magnetic halos, where
magnetic field can be estimated or measured. 
The strong magnetic fields of order of $1\mu$G are indeed
observed in galaxies and their halos, in clusters of galaxies and 
in radiolobes of radiogalaxies. As an example one can consider our
local surroundings. Milky Way belongs to the Local Group (LG)
entering the Local Supercluster (LS). LG with a size 
$\sim 1$Mpc contains 40 dwarf galaxies, two giant spirals (M31
and Milky Way) and two intermediate size galaxies. The galactic winds
cannot provide the appreciable magnetic field inside this structure. 
LS with a size of 10 -- 30 Mpc is a young system 
where dynamo mechanism cannot strengthen the  primordial magnetic field. 
In fact LS is filled by galactic clouds submerged in the voids. 
The vast majority of the
luminous galaxies reside in a small number of clouds: 98 \% of all
galaxies in 11 clouds \cite{Tully}. Thus, accepting the hypothesis
of generation of magnetic fields in compact sources, one arrives at 
the perforated  picture of the universe, with strong magnetic fields in the
compact objects and their halos (magnetic bubbles produced by galactic 
winds) and with extremely weak magnetic fields outside.  However,
even in this picture there is a scattering of UHE protons off the
magnetic bubbles and the  scattering length is 
$l_{\rm sc} \sim 1/\pi R^2 n$, where $R$ is the radius ofa  magnetic bubble
and $n$ is their space density. Among different 
 structures, the largest contribution is given by galaxy clusters which
can provide $l_{\rm sc} \sim (1-2)\times 10^3$~Mpc.  

Leaving the correlation of UHECR with AGN (BL Lacs) as an open problem
for future observations we turn now to the alternative of excluding
these correlations from analysis. The small-angle clustering then can
be probably explained in the other extreme case of very strong 
magnetic field due to lensing effect \cite{Sigl}. Rectilinear
propagation of UHE protons is not needed anymore.   

Till now we discussed mostly UHECR at $E \lsim 1\times 10^{20}$~eV. 
At higher energies, as AGASA data show, there might be an excess of
events. They should be interpreted as the new component. One
possibility is given by Superheavy Dark Matter (SDMP) \cite{SHDM}. 
The spectrum of UHE particles, produced at the SHDM decay is now reliably      
calculated using the different methods \cite{SHDMspectra}. All
calculations give very similar spectrum with $\gamma_g \approx 1.9 - 2.0$.
This spectrum is shown in Fig.1. It describes well the observed AGASA
excess. The primary particles are predicted to be UHE photons, which
at energy $E> 1\times 10^{20}$~eV cannot be excluded by available 
experimental data, in particular by the inclined Haverah Park showers
\cite{incl}\\*[1mm]

{\em In conclusion}, the observed energy spectra reveal 
the signatures of interaction of UHE protons with the CMB in the form of
the dip, beginning of the GZK cutoff and the good agreement with 
the predicted spectrum. Combined with small-angle clustering \cite{doubl}
and correlation with BL Lacs \cite{BLLac}, these data require the 
rectilinear propagation of UHE protons and AGN as their sources. 
The correlation with AGN (BL Lacs) becomes thus the most sensitive method
of measuring very weak extragalactic magnetic fields. In case this 
correlation is not confirmed by future observational data, and the 
small-angle clustering is explained by some more sophisticated phenomena
(e.g. by magnetic lensing), AGN remain the favorable UHECR sources,
but as one of the  possible candidates.\\*[1mm] 

This work was supported in part by INTAS (grant No. 99-01065). 
We are grateful to Pierre Sokolsky for active discussion and for providing
us with HiRes data. Many thanks to Masahiro Teshima for the data of AGASA,
and to M.I.Pravdin for the data of Yakutsk.
We gratefully acknowledge the useful discussions with Michael
Kachelriess and Igor Tkachev.

\widetext
\end{document}